\documentclass[11]{article}
\usepackage[utf8]{inputenc}

\usepackage{optidef}

\usepackage{pdfpages}

\usepackage{mathrsfs}
\usepackage{color,soul}
\usepackage{algorithm}
\usepackage{algpseudocode}
\usepackage{hyperref} 
\usepackage{amsmath,amssymb}
\usepackage{amsmath,amsfonts,amssymb} 
\usepackage{footnote}
\usepackage{geometry}

\usepackage{graphicx}

\title{Optimal Operation of HVDC Interconnector: Irish Case}
\author{Md Umar Hashmi}
\date{August 2022}

\begin{document}

\maketitle

\textbf{Summary}: 
In September 2018 EirGrid launched the new electricity market. These new market arrangements integrate the all island electricity market with European electricity markets, making optimal use of cross border transmission assets.
Ireland operates three operational HVDC interconnectors: Moyle, East-West, and Greenlink and one in development Celtic interconnector which connect Ireland to Scotland, Wales, and France respectively. Irish market operator, EirGrid, can maximize their operational profit by using the price difference in these electricity markets.
We propose a profit maximization modelling which considers the line losses and price difference in these different electricity markets and identifies the optimal import/export of power using HVDC interconnectors. 
These models in future should incorporate the distribution losses, renewable energy curtailment, and Irish power network congestion levels.
The proposed modeling is the first step towards implementing a multi-objective HVDC interconnector operating strategy.

\tableofcontents

\section{Motivation}
In 2018 EU energy ministers agreed an ambitious, binding renewable energy target of 32\% by 2030, up from the previous goal of 27\%. 
In Ireland and Northern Ireland, the government target of 40\% electricity consumption coming from renewable energy sources is set.
However, the Renewable Electricity Support Scheme raises this target from 40\% to 55\%.

\subsection{Flow Based Market Coupling in Europe}
\begin{figure}[!htbp]
 	\center
 	\includegraphics[width=6.5in]{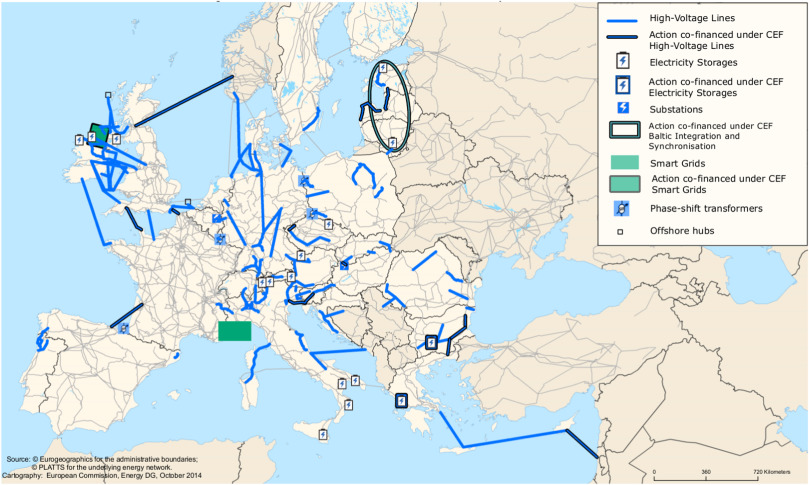}
 	\caption{\small{European HVDC interconnectors \cite{pierri2017challenges}} }\label{europehvdc}
 	\vspace{-5pt}
 \end{figure}

European Union consists of different electricity markets operating at a different frequency. Linking these electricity markets holds the potential for additional financial gains for independent utilities, increases reliability of supply, increases ability to integrate renewable energy into the power networks.
Thus, by interlinking power systems, significant welfare gains can be achieved \cite{felten2019flow}.
The European Union envisions a more interconnected electrical power system. This is evident in Figure \ref{europehvdc} showing interconnections.

With Integrated-Single Energy Market (iSEM), island of Ireland is now part of pan-European Internal Energy Market that comprises 20 countries, linked with more than 38 cross-border interconnectors, representing a total generating capacity of over 3000 terawatts \footnote{\url{http://www.eirgridgroup.com/annual-report-2018/}}.

\section{HVDC Interconnectors in Irish Power System}
There are three main sources of revenue for TSO in island of Ireland:
\begin{itemize}
	\item Transmission Use of System (TUoS): payable by all users of the transmission systems in Ireland and Northern Ireland, \vspace{-5pt}
	\item Share of tariffs as Market Operator for SEM, \vspace{-5pt}
	\item Auction receipts for the sale of capacity of the interconnectors.
\end{itemize}
In this work, we will focus on the last revenue generation by operating the HVDC interconnectors based on price potential.

\begin{figure}[!htbp]
	\center
	\includegraphics[width=5in]{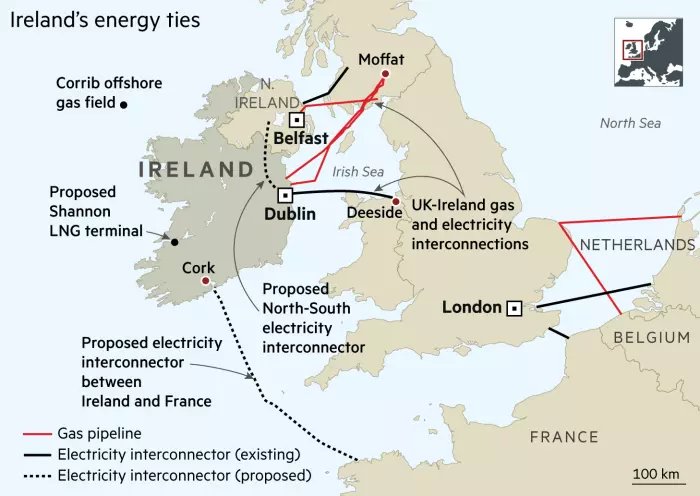}
	\caption{\small{Irish HVDC interconnectors \cite{irelandhvdc}} }\label{irelandhvdc}
	\vspace{-5pt}
\end{figure}

Figure \ref{irelandhvdc} and Table \ref{hvdcinterconnectorireland} lists the operational and ongoing HVDC interconnector projects in Ireland and Northern Ireland.

\begin{table}[!htbp]
	\small
	\caption {HVDC Interconnectors connected to island of Ireland} \label{hvdcinterconnectorireland} \vspace{-10pt}
	\begin{center}
		\begin{tabular}{| c | c| c|  c| c| }
			\hline
			Name &  Capacity &  Length & DC voltage  & Operating Since \\ 
			\hline
			HVDC Moyle Interconnector & 500 MW & 63.5 km & 250 kV & 2001 \\
			\hline
			East–West Interconnector & 500 MW  & 261 km & $\pm$ 200 kV & 2012\\
			\hline
			Greenlink Interconnector & 500 MW  & 200 km & 320kV & 2023\\
			\hline
			Celtic Interconnector & 700 MW & 575 km & $\pm$ 320 kV or $\pm$ 500 kV & 2025\\
			\hline
		\end{tabular}
		\hfill\
	\end{center}
\end{table}

\section{Optimal use of HVDC interconnectors based on price potential}

\begin{figure}[!htbp]
	\center
	\includegraphics[width=5in]{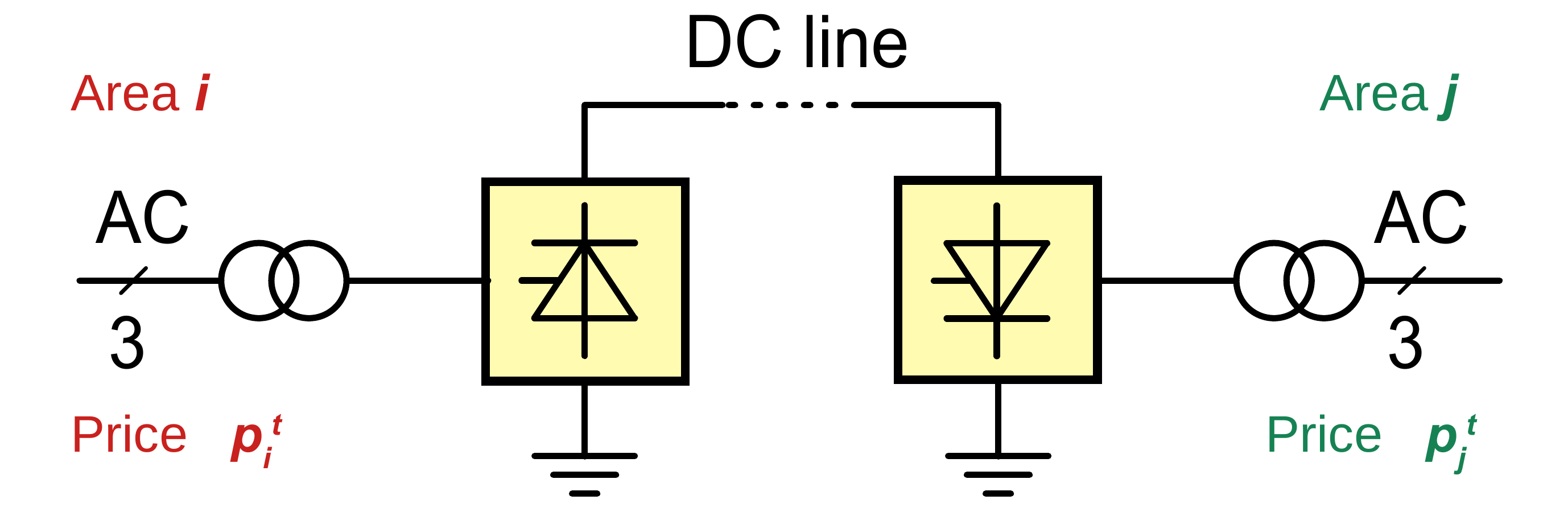}
	\caption{\small{Notation used for a 2 area network interlinked using HVDC} }\label{hvdc}
	\vspace{-5pt}
\end{figure}

Consider an HVDC interconnector linking two geographical regions $i$ and $j$ with different electricity price levels shown as $p_i^t$ and $p_j^t$ respectively, shown in Figure \ref{hvdc}.

Consider HVDC line incurs fraction $r$ as losses. Therefore, if $x$ an amount of energy is transferred at one end of HVDC link then the other end receives $(1-r)x$ an amount of energy, where $x \in [0, X_{\max}]$. $X_{\max}$ denotes the maximum power transfer capability of the HVDC link.

\textbf{Case 1}: $p_i^t > p_j^t$ then flow should be from region $j$ to $i$. Here $t$ denotes time. The condition for operating the HVDC link would be
\begin{equation}
\frac{p_i^t}{p_j^t} > \frac{1}{1-r}
\label{equationcond1}
\end{equation}
In case Eq.~\ref{equationcond1} is true then the profit will be strictly positive, given as
\begin{gather*}
\text{Profit}_{C1} = p_i^t (1-r) x - p_j^tx >0.
\end{gather*}
This can be given as
\begin{equation}
\text{Profit}_{C1} = \max\Big( \{(p_i^t - p_j^t) - r p_i^t \}x,0  \Big)
\end{equation}

\textbf{Case 2}: $p_j^t > p_i^t$ then flow should be from region $i$ to $j$. The condition for operating the HVDC link would be
\begin{equation}
\frac{p_j^t}{p_i^t} > \frac{1}{1-r}
\label{equationcond2}
\end{equation}
In case Eq.~\ref{equationcond2} is true then the profit will be strictly positive, given as
\begin{gather*}
\text{Profit}_{C2} = p_j^t (1-r) x - p_i^tx >0.
\end{gather*}
This can be given as
\begin{equation}
\text{Profit}_{C2} = \max\Big( \{(p_j^t - p_i^t) - r p_j^t \}x,0  \Big)
\end{equation}

Both the cases can be combined to calculate total profit as
\begin{equation}
\text{Profit} = \text{Profit}_{C1}+ \text{Profit}_{C2} = x \max\Big(p_i^t - p_j^t - r p_i^t, p_j^t - p_i^t - r p_j^t,0  \Big)
\label{totalprofit}
\end{equation}

The total profit shown in Eq. \ref{totalprofit}. Operating an HVDC link for even a small profit can lead to bang-bang control of the link. For eliminating low returning transactions, a bias needs to be created \cite{hashmi2018limiting}.

\begin{equation}
\text{Profit} = \text{Profit}_{C1}+ \text{Profit}_{C2} = x \max\Big(p_i^t - p_j^t - r p_i^t - R_b, p_j^t - p_i^t - r p_j^t - R_b,0  \Big)
\label{totalprofit2}
\end{equation}
here $R_b>0$ denotes bias term.

\subsection{Formulations considering network limitations}
The power network limitations will constraint the maximum import and export possible from the HVDC link. This can be denoted as $x_t \in [0, X_{\max}^t]$. $X_{\max}^t$ is dynamic and denotes capability of the network to import and export.
\begin{mini!}[1]
	{x_1, \dots, x_{T}}{ \sum_{i=1}^{T} x_t \lambda_t}{}{(P)}
	\label{eq:optprob}
	\addConstraint{x_{t} }{\in [0, X_{\max}^t]}{ \ t=1, \ldots,T}
	\addConstraint{\lambda_t }{\geq 0 }{t=1, \ldots,T}
	\addConstraint{\lambda_t }{\geq p_i^t - p_j^t - r p_i^t }{t=1, \ldots,T}
	\addConstraint{\lambda_t }{\geq p_j^t - p_i^t - r p_j^t }{t=1, \ldots,T.}
\end{mini!}

\subsection{Wheeling criteria for HVDC interconnectors: economic perspectives}

\begin{figure}[!htbp]
	\center
	\includegraphics[width=4in]{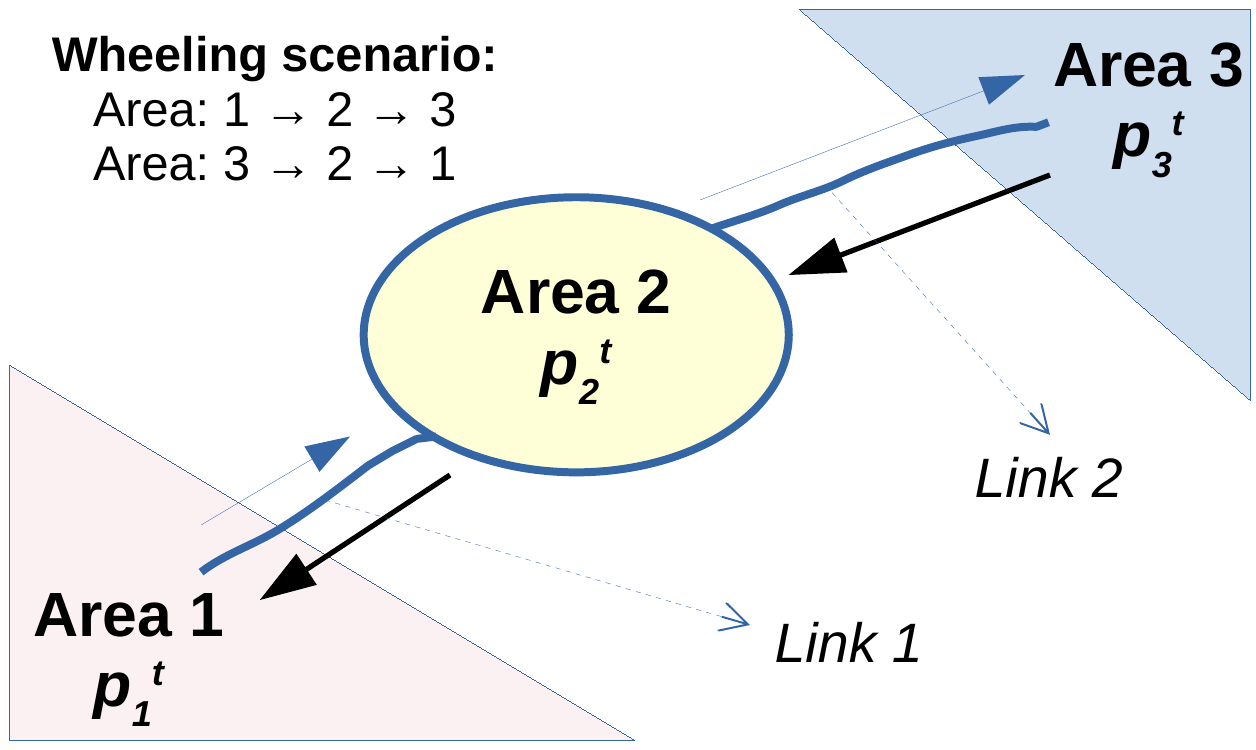}
	\caption{\small{Wheeling scenario} }\label{wheel}
	\vspace{-5pt}
\end{figure}

Consider a 3 area which are interconnected through HVDC links. The stylized diagram is shown in Figure \ref{wheel}. Wheeling refers to flow of power from Area 1 to 3 via area 2 or from area 3 to 1 via 2.

Consider losses in Link 1 as $r_1$ and Link 2 as $r_2$.\\

\textbf{Scenario 1}: Flow of energy from area 1 to 2 to 3

Energy transferred from area 1 is $x$. Energy reaching region 2 is $x(1-r_1)$ and region 3 is $x(1-r_1)(1-r_2)(1-c)$, where $c$ denotes transmission losses in Area 2. The criteria for wheeling from 1 to 3 are
\begin{gather}
p_3^t(1-r_2)(1-c) - p_2^t > 0,\\
p_2^t(1-r_1) - p_1^t > 0.
\end{gather}

The operational profit for this scenario is given as
\begin{equation}
\text{Profit}_{123}^t = \Big( p_3^t (1-r_1)(1-r_2)(1-c) - p_1^t \Big) x.
\end{equation}

\textbf{Scenario 2}: Flow of energy from area 3 to 2 to 1

Energy transferred from area 3 is $x$. Energy reaching region 2 is $x(1-r_2)$ and region 1 is $x(1-r_1)(1-r_2)(1-c)$, where $c$ denotes transmission losses in Area 2. The criteria for wheeling from 1 to 3 are
\begin{gather}
p_1^t(1-r_1)(1-c) - p_2^t > 0,\\
p_2^t(1-r_2) - p_3^t > 0.
\end{gather}

The operational profit for this scenario is given as
\begin{equation}
\text{Profit}_{321}^t = \Big( p_1^t (1-r_1)(1-r_2)(1-c) - p_3^t \Big) x.
\end{equation}

\section{Numerical results}
We present a case-study for Irish power system consisting of HVDC links, tabulated in Table \ref{hvdcinterconnectorireland}.
Assumptions for the numerical simulation are as follows
\begin{itemize}
	\item Island of Ireland have a single electricity price,
	\item HVDC interconnectors incurs 1\% loss per 100 km,
\end{itemize}

\begin{table}[!htbp]
	\small
	\caption {Parameters for HVDC Interconnectors connected to island of Ireland} \label{hvdcnumericalireland} \vspace{-10pt}
	\begin{center}
		\begin{tabular}{| c | c| c|   }
			\hline
			Name &  Capacity ($X_{\max}$)  &Loss fraction (r) \\ 
			\hline
			Moyle (Ireland - Scotland) & 500 MW &  0.635/100 \\
			\hline
			East–West Interconnector (Ireland - Wales) & 500 MW  &  2.61/100\\
			\hline
			Greenlink (Ireland - Wales) & 500 MW  &  2/100\\
			\hline
			Celtic Interconnector (Ireland - France) & 700 MW & 5.75/100\\
			\hline
		\end{tabular}
		\hfill\
	\end{center}
\end{table}

Consider the following electricity prices for the numerical simulation shown in Table \ref{electricityprices}.

\begin{table}[!htbp]
	\small
	\caption {Electricity prices for numerical simulation} \label{electricityprices} \vspace{-10pt}
	\begin{center}
		\begin{tabular}{| c | c|   }
			\hline
			Region  &  Price level (euros/MWh)   \\ 
			\hline
			Ireland  & 100 \\
			\hline
			Scotland & 120 \\
			\hline
			 Wales & 75 \\
			\hline
			France & 50 \\
			\hline
		\end{tabular}
		\hfill\
	\end{center}
\end{table}

The profit in using Celtic at the price levels listed in Table \ref{electricityprices} and parameters listed in Table \ref{hvdcnumericalireland} is found to be 30975 euros. The profit for usinf EWI and Greenlink is 11195 euros respectively and Moyle is 9622 euros. Thus the total profit by operating all Irish interconnectors are found to be 61414 euros. Note that the profit is by operating the interconnector in just one hour. A linear extrapolation would imply operating income exceeding 525 million euros in 1 year.

\section{Conclusion}
This paper proposes optimal operation of HVDC links based on price potential. The model also considers losses in these links.
Numerical results indicate that these links can generate significant financial opportunities by facilitating a more interconnected power network, which the European Union aspires for.

\bibliographystyle{IEEEtran}
\bibliography{referencesDoc}


\end{document}